\pdfoutput=1

\documentclass[twocolumn,showpacs,amsmath,amssymb,prl]{revtex4}

\usepackage{graphicx}
\usepackage{dcolumn}
\usepackage{bm}
\usepackage{epsfig}
\usepackage{amsmath}
\usepackage{color}

\newcommand{\be}{\begin{equation}}
\newcommand{\ee}{\end{equation}}
\newcommand{\ba}{\begin{eqnarray}}
\newcommand{\ea}{\end{eqnarray}}

\begin{document}
\input{epsf}

\title{High Energy Neutrinos from the Gravitational Wave event GW150914 possibly associated with a short Gamma-Ray Burst}

\author{Reetanjali Moharana$^1$}
\email{reetanjalim@uj.ac.za}

\author{Soebur Razzaque$^1$}
\email{srazzaque@uj.ac.za}

\author{Nayantara Gupta$^2$}
\author{Peter M\'esz\'aros$^3$}

\affiliation{$^1$Department of Physics, University of Johannesburg, PO Box 524, Auckland Park 2006, South Africa}

\affiliation{$^2$Raman Research Institute, Sadashivanagar, Bangalore 560080, India}

\affiliation{$^3$Center for Particle and Gravitational Astrophysics,
Dept.\ of Astronomy \& Astrophysics and Dept.\ of Physics,
Pennsylvania State University, University Park, PA 16802, USA}

\begin{abstract}
High-energy neutrino (HEN) and gravitational wave (GW) can probe astrophysical sources in addition to electromagnetic observations. Multimessenger studies can reveal nature of the sources which may not be discerned from one type of signal alone. We discuss HEN emission in connection with the Advanced Laser Interferometer Gravitational-wave Observatory (ALIGO) event GW150914 which could be associated with a short gamma-ray burst (GRB) detected by the {\it Fermi} Gamma-ray Burst Monitor (GBM) 0.4~s after the GW event and within localization uncertainty of the GW event. We calculate HEN flux from this short GRB, GW150914-GBM, and show that non-detection of a high-energy starting event (HESE) by the IceCube Neutrino Observatory can constrain the total isotropic-equivalent jet energy of this short burst to be less than $3\times 10^{52}$~erg.  
\end{abstract}


\date{\today}
\maketitle

\section{Introduction}

The detection of GW150914 on September 14, 2015 at 09:50:45 UTC is a watershed event~\cite{Abbott:2016blz}. A sweeping up of the GW frequency and subsequent ringdown heralded formation of a black hole from a binary merger in a split second. GW has long been hypothesized as a cosmic messenger and detection of GW150914 has opened a new window to the universe. A few years earlier the IceCube Neutrino Observatory detected cosmic high-energy neutrinos for the first time~\cite{Aartsen:2013bka} and ushered an era of multimessenger astronomy. 

Binary mergers of neutron stars (NSs) or black hole (BH) and NS systems have been proposed to be the progenitors of short GRBs~\cite{Eichler:1989ve,Narayan:1992iy}. Since GW is naturally produced in binary mergers, it has been predicted to be coincident with the short GRBs~\cite{Eichler:1989ve,Kobayashi:2002by,Rezzolla:2011da}. Detection of a short GRB of duration 1~s, GW150914-GBM within 0.4~s and from within $\sim 600$ square degree arrival direction uncertainty of the GW150914 by the {\it Fermi}-GBM~\cite{Connaughton:2016umz} is therefore very intriguing. GW150914-GBM could not be characterized very well due to large uncertainty in the {\it Fermi}-GBM position, (RA, Dec) = ($57^\circ$, $-22^\circ$) with a $68\%$ uncertainty region over 9000 square degrees in the standard 50-300~keV analysis; (RA, Dec) = ($75^\circ$, $-73^\circ$) with a $68\%$ uncertainty region about 3000 square degrees for 100-1000~keV analysis. GW150914-GBM was outside of the field of view of the {\it Fermi} Large Area Telescope (LAT) initially and no GeV afterglow was detected when it could observe~\cite{Fermi-LAT:2016qqr}. A joint LIGO-{\it Fermi} analysis reduces the $90\%$ confidence region for GW150914 to 199 square degrees and located in the Southern hemisphere (see, Fig.~8 bottom right panel in Ref.~\cite{Connaughton:2016umz}). The INTEGRAL/SPI-ACS also observed the GW150914 localization region in hard X-ray without detection~\cite{Savchenko:2016kiv}.

The isotropic-equivalent luminosity of GW150914-GBM in the 1~keV to 10 MeV energy range was $1.8^{+1.5}_{-1.0} \times 10^{49}$~erg/s~\cite{Connaughton:2016umz} using a $410^{+160}_{-180}$~Mpc distance inferred from the GW150914 event~\cite{Abbott:2016blz}. The true luminosity and hence the true energy radiated, depending on the GRB jet opening angle, is much lower. As for comparison, the total energy radiated in GW was $(5.4\pm 0.9)\times 10^{54}$~erg~\cite{Abbott:2016blz}.

A binary merger of two BHs, as proposed for the origin of the  GW150914 event~\cite{Abbott:2016blz}, is traditionally thought to produce no electromagnetic counterpart due to a lack of accretion material. Detection of GW150914-GBM~\cite{Connaughton:2016umz}, however, have raised the possibility that the binary BH systems may also possess some accretion material. In a recent work, the authors of Ref.~\cite{Perna:2016jqh} modeled that one of the BHs of the binary system could possess a long-lived dormant accretion disk created from the explosion of its low-metallicity progenitor star. This disk is fully revived only during the final merger of the BHs and is accreted rapidly to power a short GRB. In the particular case of GW150914, an accretion disk of mass $\sim (10^{-4} - 10^{-3})M_{\odot}$ around a $\sim (30-40)M_\odot$ BH may exist~\cite{Perna:2016jqh}, which is sufficient to power GW150914-GBM with a significant baryon load in its jet.

High energy neutrinos can be produced in the GRB jet from interactions of shock-accelerated protons with the observed $\gamma$ rays~\cite{Waxman:1997ti, Dermer:2003zv, Razzaque:2003uw, Guetta:2003wi, Murase:2005hy, Moharana:2011hh, Hummer:2011ms}. Detection of HEN from GW150914-GBM could potentially measure the total jet power and yet uncertain jet bulk Lorentz factor for the short GRBs. The ANTARES and IceCube Neutrino Observatories have searched for HEN in coincidence with the GW150914 but no detection has been reported~\cite{Adrian-Martinez:2016xgn}.

In this work we calculate the HEN flux using the short GRB characteristics of GW150914-GBM and estimate the corresponding number of HESE events at the IceCube detector. Using the fact that no HEN event was detected, we put constraints on the power and bulk Lorentz factor of the short GRB jet.

\section{Neutrino flux and HESE}

We calculate the HEN flux from the short GRB (duration $\sim 1$~s) detected by the {\it Fermi} GBM~\cite{Connaughton:2016umz} in-coincidence with the GW event. The isotropic-equivalent electromagnetic luminosity of the short burst is $L_\gamma = 1.8\times 10^{49}$~erg/s~\cite{Connaughton:2016umz}. We assume protons, accelerated in the internal shocks, interact with these photons ($p\gamma$) to produce neutrinos~\cite{Waxman:1997ti} and calculate their flux following Refs.~\cite{Gupta:2006jm,Moharana:2011hh}.

The target photon density for $p\gamma$ interactions can be written in terms of a broken power-law with indices $\alpha$, $\beta$ and break energy $\epsilon_{\gamma, b}$ in the GRB jet frame as 
\begin{equation}
\frac{dn_{\gamma}}{d\epsilon_{\gamma}}
=A \left\{ \begin{array}{l@{\quad \quad}l}
\epsilon_{\gamma}^{-\alpha}; &
\epsilon_{\gamma}<\epsilon_{\gamma,b}\\
{\epsilon_{\gamma,b}}^{\beta-\alpha}
\epsilon_{\gamma}^{-\beta}; & \epsilon_{\gamma}>\epsilon_{\gamma,b}
\end{array}\right. .
\label{photon}
\end{equation} 
The normalization constant $A$ is given by,
\begin{equation}
A=\frac{U_{\gamma}{\epsilon_{\gamma,b}}^{\alpha-2}}{[\frac{1}{\beta-2}-\frac{1} {\alpha-2}]},
\label{A_g}
\end{equation}
where $U_{\gamma}=4 \pi D_L ^2 F_{\gamma}/(4\pi R^2 \Gamma^2 c)$ is the internal energy density in photons with flux $F_{\gamma}$, luminosity distance $D_L$, jet radius $R$ and bulk Lorentz factor $\Gamma$. The efficiency for pion production, with a fractional energy transfer from a proton to a pion from $p\gamma$ interaction can be written, using Eq.~(\ref{A_g}), as
\begin{equation}
f_{\pi} (\epsilon_p)= f_0^{\pi} \left\{\begin{array}{l@{\quad\quad}l}
\frac{1.34^{\alpha-1}}{\alpha+1}\left(\frac{\epsilon_p}{\epsilon_{p,b}}\right)^{\alpha-1}; & {\epsilon_p>\epsilon_{p,b}}\\
\frac{1.34^{\beta-1}}{\beta+1}\left(\frac{\epsilon_p}{\epsilon_{p, b}}\right)^{\beta-1}; & {\epsilon_p < \epsilon_{p,b}}
\end{array}
\right. .
\label{fpi}
\end{equation}
Here the prefactor is given by
\begin{equation}
f_0^{\pi}=\xi_{\pi}\frac{4.5L_{\gamma,51}}{\Gamma_{300}^4 \,
t_{v,-3} (\epsilon_{\gamma,b}/{\rm MeV})}\frac{1}{\big
[\frac{1}{\beta-2}-\frac{1}{\alpha-2}\big ]},
\end{equation}
where $L_\gamma = 10^{51}L_{\gamma,51}$~erg/s, $\Gamma = 300 \Gamma_{300}$ and $t_v = 10^{-3}t_{v,-3}$~s is the flux variability time.  The factor $\xi_\pi=0.2$ is the average fraction of energy lost to the pions. The proton break energy in Eq.~(\ref{fpi}), corresponding to the photon break energy $\epsilon_{\gamma,b}$ in Eq.~(\ref{A_g}), is  $\epsilon_{p,b} = 1.3\times 10^{7} \Gamma_{300}^2(\epsilon_{\gamma,b}/{\rm MeV})^{-1}$~GeV. The pion production efficiency in Eq.~(\ref{fpi}) is restricted as $f_{\pi} \le 1$. Note that the jet radius at which emission takes place is $R = \Gamma^2 ct_v$.  

We calculate the neutrino flux in this $p\gamma\to \Delta$-resonance channel for a $\epsilon_p^{-2}$ proton spectrum. The HEN flux for one neutrino flavor from the pion decay ($\pi^+ \to \mu^+ + \nu_\mu$) or muon decay ($\mu^+ \to e^+ + \nu_e + {\bar \nu}_\mu$) is given by
\begin{equation}
\epsilon_{\nu}^2\frac{dN_{\nu}}{d\epsilon_{\nu}}\approx \frac{f_{\pi}}
{8\kappa} \frac{\eta_pL_{\gamma}}{4\pi D_L^2}
\left\{\begin{array}{l} 1; \hspace{1.5cm} \epsilon_{\nu}<\epsilon_{\nu,s}\\
 (\frac{\epsilon_{\nu}}{\epsilon_{\nu,s}})^{-2}; \hspace{0.5cm} \epsilon_{\nu}>\epsilon_{\nu,s}.
\end{array} \right.,
\label{totnupi}
\end{equation}
where $\eta_pL_{\gamma}$ is the energy in protons, $\kappa = 1.8$ is a normalization factor since $L_\gamma$ is the bolometric luminosity in the 1~keV to 10 ~MeV range \cite{Gupta:2006jm}. In Eq.~(\ref{totnupi}) $\epsilon_{\nu,s}$ is a break energy arising from synchrotron cooling of pions and muons in the magnetic field of the jet~\cite{Rachen:1998fd}. We assume the energy density in the magnetic field is $B^2/8\pi = U_\gamma$. We use this magnetic field to calculate the pion and muon synchrotron break energies by equating their respective synchrotron cooling time scales with the dynamic time scale $\Gamma t_v$ in the jet frame. We also calculate the maximum accelerated proton energy in this magnetic field which is limited by proton cooling time scales and/or the jet dynamic time scale. See also Ref.~\cite{Petropoulou:2014awa} for the role of $\eta_p$ in case of proton acceleration.

\begin{figure}[tbp]
\centering 
\includegraphics[width=20pc]{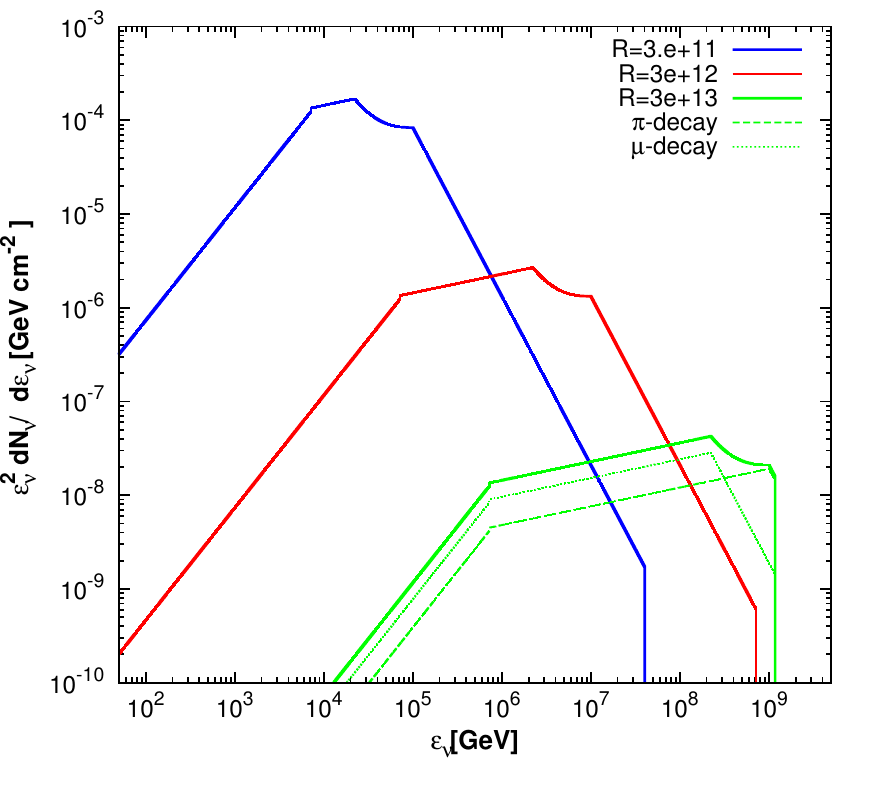}
\caption{\label{flux} {Total (3 flavor) neutrino flux (blue, red and green solid lines) from pion and muon decays in the internal shocks for different values of $R = \Gamma^2 ct_v$ in cm with different values of $\Gamma$ and with $t_v = 10^{-2}$~s. We kept fixed $\eta_p=1$ in Eq.~(\ref{totnupi}). The dashed and dotted green lines represent the pion- and muon-decay components for one case. The flux models are calculated following Refs.~\cite{Gupta:2006jm, Moharana:2011hh}. Also shown in gray lines are conventional atmospheric neutrino fluxes~\cite{Honda:2006qj} for two different zenith angles $\theta_z$.}}
\end{figure}

Figure~\ref{flux} shows the neutrino fluxes for different model parameters, considering GW150914-GBM as a short GRB. Here we have used $\alpha= 1.2$, $\beta= 2.2$ and $\epsilon_{\gamma,b} = 1$~MeV, $L_\gamma = 1.8\times 10^{49}$~erg/sec and $\eta_p = 1$. We have plotted the HEN flux for three different jet radius $R$ by varying $\Gamma$ while keeping fixed $t_v = 10^{-2}$~s. Although the values of the parameters $\alpha$, $\beta$, $\epsilon_{\gamma,b}$ and $t_v$ are not well-constrained from the {\it Fermi} GBM observations of the GW150914-GBM~\cite{Connaughton:2016umz}, they are within typical ranges of their values for short GRBs. Figure~\ref{flux} also shows conventional atmospheric neutrino flux~\cite{Honda:2006qj} which has been measured by IceCube up to 400 TeV~\cite{Abbasi:2010ie}.

Finally we calculate the HESE number (cascade or track) at the IceCube detector from GW150914-GBM using the corresponding neutrino effective area $A_{\nu, {\rm eff}} (\epsilon_\nu)$, averaged over the full sky in the $\sim 25$~TeV - 1~PeV energy range~\cite{Aartsen:2013jdh}, as
\begin{equation}
N_\nu = T \int_{25~{\rm TeV}}^{1~{\rm PeV}} {\frac{dN_{\nu}}{d\epsilon_{\nu}} A_{\nu,{\rm eff}}}(\epsilon_\nu) d\epsilon_\nu,
\label{nuneu}
\end{equation}
where $T = 1$~s. Note that the $90\%$ confidence region of the joint LIGO-{\it Fermi} localization of the GW150914 event (bottom right panel of Fig.~8 in Ref.~\cite{Connaughton:2016umz}) lies entirely in the Southern hemisphere at $Dec < -60^\circ$. Neutrinos with $\sim 1$ PeV energy in the upper limit of integration in Eq.~(\ref{nuneu}) suffers negligible absorption inside the earth for this source localization region. The atmospheric neutrino flux is much lower than the flux from the GW150914-GBM in this energy range for $R \lesssim 3\times 10^{12}$~cm (see Fig.~\ref{flux}).

\section{Results and Discussion}

\begin{figure}[tbp]
\centering 
\includegraphics[width=21pc]{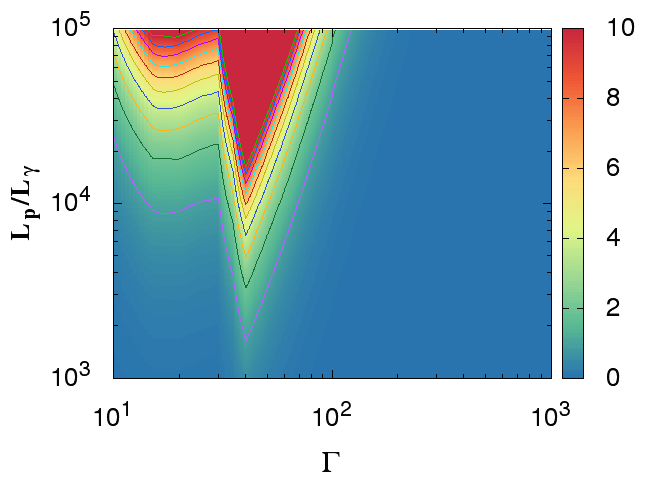}
\includegraphics[width=21pc]{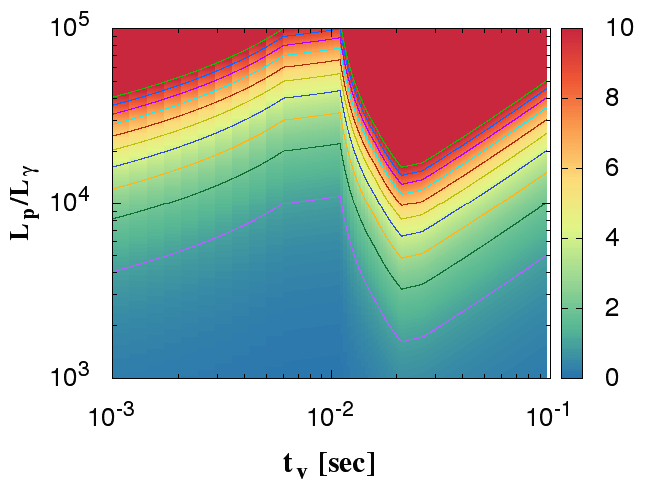}
\caption{
\label{hese} {Number of high-energy starting events in the IceCube detector calculated from the flux of the short GRB in coincidence with GW150914. The ratio of the proton to electromagnetic energy $\eta_p = L_p/L_\gamma$ is plotted against the jet bulk Lorentz factor $\Gamma$ (top panel) and the flux variability time $t_v$ (bottom panel). The contour lines represent 1, 2, 3 ... etc.\ HESE (from the lowest and above). The minimum values of $\eta_p$, $\Gamma$ and $t_v$ for detection of 1 HESE are $1.6\times 10^3$, $37$ and $2\times 10^{-2}$~s, respectively. Non-detection of any event results in a limit on the isotropic-equivalent jet power $L_p \lesssim 3\times 10^{52}$~erg/s.}}
\end{figure}

Our main results are shown in Fig.~\ref{hese} where we have plotted the number of HESE (cascade- or track-type) in IceCube from the short burst GW150914-GBM in coincidence with the gravitational wave detection. In the top panel we plot the bulk Lorentz factor $\Gamma$ of the GRB jet in the $x$-axis, which is the most-sensitive parameter for calculating the $p\gamma$ interaction efficiency in Eq.~(\ref{fpi}). We have kept the flux variability time $t_v = 10^{-2}$~s fixed. In the bottom panel we plot the flux variability time $t_v$ in the $x$-axis, which is the other sensitive parameter. We have kept $\Gamma = 10^{1.5}$ fixed in this case. The $y$-axis of Fig.~\ref{hese} (both panels) shows the ratio of the proton to photon energy $\eta_p = L_p/L_\gamma$ in Eq.~(\ref{totnupi}). Different shading in the plot represents different number of events as indicated in the sidebars. The contour lines indicate number of HESE as 1, 2, 3, etc.\ from the bottom and above. 

Note in Fig.~\ref{hese} (top panel) that the event number is lower for higher $\Gamma \gtrsim 10^2$, as expected, and deduced from non-detection of GRBs in neutrinos~\cite{Ahlers:2011jj, He:2012tq, Gao:2013fra}. For $\Gamma \lesssim 10$, the peak in Fig.~\ref{flux} shifts to further below the 25~TeV threshold energy for HESE detection. Thus the HESE detection is the most effective for $10 \lesssim \Gamma \lesssim 10^2$. The $t_v$ dependence (bottom panel) is milder than the $\Gamma$ dependence, with a preference for tens of ms variability. The preferred range of radii for HEN production is therefore between $\sim 3\times 10^{10}$~cm and $\sim 3\times 10^{12}$~cm. The atmospheric neutrino flux is lower than the HEN flux in the energy range of Eq.~(\ref{nuneu}) for these radii even in case of $\eta_p = 1$ (see Fig.~\ref{flux}).

The plots in Fig.~\ref{hese} allow us to put constraints on the $\eta_p - \Gamma$ and $\eta_p - t_v$ parameter spaces using non-detection of any HEN by IceCube in coincidence with the GW event~\cite{Adrian-Martinez:2016xgn}. The minimum parameter values for detection of 1 HESE is $(\eta_p,~\Gamma) = (1.6\times 10^3,~37)$ (top panel) and $(\eta_p,~t_v) = (1.6\times 10^3,~2.1\times 10^{-2}~{\rm s})$ (bottom panel). The $t_v = 10^{-2}$ value kept fixed in the top panel is largely consistent with the value in the bottom panel where $\eta_p$ is the minimum. Similarly the $\Gamma=10^{1.5}$ value kept fixed in the bottom panel is largely consistent with the value obtained in the top panel where $\eta_p$ is the minimum. As a result the minimum $\eta_p$ values are also consistent with each other. A detection of at least 1 HESE would then require $L_p/L_\gamma \gtrsim 1.6\times 10^3$ or $L_p \sim 3\times 10^{52}$~erg/s, at least. Given that the short GRB had a duration of $\sim 1$~s, the upper limit on the isotropic-equivalent proton energy is then $\sim 3\times 10^{52}$~erg. Because of the relativistic beaming effect the true GRB jet energy must be smaller by roughly two orders of magnitude.  This implies that $\ll 0.5\%$ of the energy was emitted as kinetic energy as compared to the GW energy.  

High-energy neutrino emission from gravitational wave sources has been discussed recently~\cite{Razzaque:2006ju, Murase:2006mm, Wang:2007ya, Murase:2009pg, Bartos:2011aa}. Short GRBs can potentially be the multimessenger sources of electromagnetic, neutrino and gravitational wave. Detection of 100 GeV $\gamma$ rays simultaneously with neutrinos from short GRBs~\cite{Razzaque:2006ju} might be possible as well in near future with sensitive ground-based Cherenkov $\gamma$-ray detector such as the High Altitude Water Cherenkov (HAWC) Observatory~\cite{Abeysekara:2011yu}. 

\par

We thank Kohta Murase and Maria Petropoulou for useful comments. This work was supported in part by the National Research Foundation (South Africa) grants no.\ 87823 (CPRR) and no.\ 93273 (MWGR) to SR. PM acknowledges support from a NASA grant NNX13AH50G.


\begin{thebibliography}{srt}


\bibitem{Abbott:2016blz} 
  B.~P.~Abbott {\it et al.} [LIGO Scientific and Virgo Collaborations],
  Phys.\ Rev.\ Lett.\  {\bf 116}, no. 6, 061102 (2016)
  doi:10.1103/PhysRevLett.116.061102
  [arXiv:1602.03837 [gr-qc]].

\bibitem{Aartsen:2013bka} 
  M.~G.~Aartsen {\it et al.}  [IceCube Collaboration],
  Phys.\ Rev.\ Lett.\  {\bf 111}, 021103 (2013)
  [arXiv:1304.5356 [astro-ph.HE]].

\bibitem{Eichler:1989ve} 
  D.~Eichler, M.~Livio, T.~Piran and D.~N.~Schramm,
  Nature {\bf 340}, 126 (1989).
  doi:10.1038/340126a0

\bibitem{Narayan:1992iy} 
  R.~Narayan, B.~Paczynski and T.~Piran,
  Astrophys.\ J.\  {\bf 395}, L83 (1992)
  doi:10.1086/186493
  [astro-ph/9204001].

\bibitem{Kobayashi:2002by} 
  S.~Kobayashi and P.~Meszaros,
  Astrophys.\ J.\  {\bf 589}, 861 (2003)
  doi:10.1086/374733
  [astro-ph/0210211].

\bibitem{Rezzolla:2011da} 
  L.~Rezzolla, B.~Giacomazzo, L.~Baiotti, J.~Granot, C.~Kouveliotou and M.~A.~Aloy,
  Astrophys.\ J.\  {\bf 732}, L6 (2011)
  doi:10.1088/2041-8205/732/1/L6
  [arXiv:1101.4298 [astro-ph.HE]].

\bibitem{Connaughton:2016umz} 
  V.~Connaughton {\it et al.},
  arXiv:1602.03920 [astro-ph.HE].

\bibitem{Fermi-LAT:2016qqr} 
  [Fermi-LAT Collaboration],
  arXiv:1602.04488 [astro-ph.HE].

\bibitem{Savchenko:2016kiv} 
  V.~Savchenko {\it et al.},
  arXiv:1602.04180 [astro-ph.HE].

\bibitem{Perna:2016jqh} 
  R.~Perna, D.~Lazzati and B.~Giacomazzo,
  Astrophys.\ J.\ Lett.\ (accepted) [arXiv:1602.05140 [astro-ph.HE]].

\bibitem{Waxman:1997ti} 
  E.~Waxman and J.~N.~Bahcall,
  Phys.\ Rev.\ Lett.\  {\bf 78}, 2292 (1997)
  doi:10.1103/PhysRevLett.78.2292
  [astro-ph/9701231].

\bibitem{Dermer:2003zv} 
  C.~D.~Dermer and A.~Atoyan,
  Phys.\ Rev.\ Lett.\  {\bf 91}, 071102 (2003)
  doi:10.1103/PhysRevLett.91.071102
  [astro-ph/0301030].

\bibitem{Razzaque:2003uw} 
  S.~Razzaque, P.~Meszaros and E.~Waxman,
  Phys.\ Rev.\ D {\bf 69}, 023001 (2004)
  doi:10.1103/PhysRevD.69.023001
  [astro-ph/0308239].

\bibitem{Guetta:2003wi} 
  D.~Guetta, D.~Hooper, J.~Alvarez-Muniz, F.~Halzen and E.~Reuveni,
  Astropart.\ Phys.\  {\bf 20}, 429 (2004)
  doi:10.1016/S0927-6505(03)00211-1
  [astro-ph/0302524].

\bibitem{Murase:2005hy} 
  K.~Murase and S.~Nagataki,
  Phys.\ Rev.\ D {\bf 73}, 063002 (2006)
  doi:10.1103/PhysRevD.73.063002
  [astro-ph/0512275].
  
\bibitem{Gupta:2006jm} 
  N.~Gupta and B.~Zhang,
  Astropart.\ Phys.\  {\bf 27}, 386 (2007)
  doi:10.1016/j.astropartphys.2007.01.004
  [astro-ph/0606744].

\bibitem{Moharana:2011hh} 
  R.~Moharana and N.~Gupta,
  Astropart.\ Phys.\  {\bf 36}, 195 (2012)
  doi:10.1016/j.astropartphys.2012.05.019
  [arXiv:1107.4483 [astro-ph.HE]].

\bibitem{Rachen:1998fd} 
  J.~P.~Rachen and P.~Meszaros,
  Phys.\ Rev.\ D {\bf 58}, 123005 (1998)
  doi:10.1103/PhysRevD.58.123005
  [astro-ph/9802280].

\bibitem{Hummer:2011ms} 
  S.~Hummer, P.~Baerwald and W.~Winter,
  Phys.\ Rev.\ Lett.\  {\bf 108}, 231101 (2012)
  doi:10.1103/PhysRevLett.108.231101
  [arXiv:1112.1076 [astro-ph.HE]].

\bibitem{Adrian-Martinez:2016xgn} 
  S.~Adrian-Martinez {\it et al.} [ANTARES and IceCube and LIGO Scientific and Virgo Collaborations],
  arXiv:1602.05411 [astro-ph.HE].

\bibitem{Petropoulou:2014awa} 
  M.~Petropoulou,
  Mon.\ Not.\ Roy.\ Astron.\ Soc.\  {\bf 442}, no. 4, 3026 (2014)
  doi:10.1093/mnras/stu1079
  [arXiv:1405.7669 [astro-ph.HE]].

\bibitem{Honda:2006qj} 
  M.~Honda, T.~Kajita, K.~Kasahara, S.~Midorikawa and T.~Sanuki,
  Phys.\ Rev.\ D {\bf 75}, 043006 (2007)
  doi:10.1103/PhysRevD.75.043006
  [astro-ph/0611418].

\bibitem{Abbasi:2010ie} 
  R.~Abbasi {\it et al.} [IceCube Collaboration],
  Phys.\ Rev.\ D {\bf 83}, 012001 (2011)
  doi:10.1103/PhysRevD.83.012001
  [arXiv:1010.3980 [astro-ph.HE]].

\bibitem{Aartsen:2013jdh} 
  M.~G.~Aartsen {\it et al.} [IceCube Collaboration],
  Science {\bf 342}, 1242856 (2013)
  doi:10.1126/science.1242856
  [arXiv:1311.5238 [astro-ph.HE]].

\bibitem{Ahlers:2011jj} 
  M.~Ahlers, M.~C.~Gonzalez-Garcia and F.~Halzen,
  Astropart.\ Phys.\  {\bf 35}, 87 (2011)
  doi:10.1016/j.astropartphys.2011.05.008
  [arXiv:1103.3421 [astro-ph.HE]].

\bibitem{He:2012tq} 
  H.~N.~He, R.~Y.~Liu, X.~Y.~Wang, S.~Nagataki, K.~Murase and Z.~G.~Dai,
  Astrophys.\ J.\  {\bf 752}, 29 (2012)
  doi:10.1088/0004-637X/752/1/29
  [arXiv:1204.0857 [astro-ph.HE]].

\bibitem{Gao:2013fra} 
  S.~Gao, K.~Kashiyama and P.~Meszaros,
  Astrophys.\ J.\  {\bf 772}, L4 (2013)
  doi:10.1088/2041-8205/772/1/L4
  [arXiv:1305.6055 [astro-ph.HE]].

\bibitem{Razzaque:2006ju} 
  S.~Razzaque and P.~Meszaros,
  Astrophys.\ J.\  {\bf 650}, 998 (2006)
  doi:10.1086/507261
  [astro-ph/0601652].

\bibitem{Murase:2006mm} 
  K.~Murase, K.~Ioka, S.~Nagataki and T.~Nakamura,
  Astrophys.\ J.\  {\bf 651}, L5 (2006)
  doi:10.1086/509323
  [astro-ph/0607104].

\bibitem{Wang:2007ya} 
  X.~Y.~Wang, S.~Razzaque, P.~Meszaros and Z.~G.~Dai,
  Phys.\ Rev.\ D {\bf 76}, 083009 (2007)
  doi:10.1103/PhysRevD.76.083009
  [arXiv:0705.0027 [astro-ph]].

\bibitem{Murase:2009pg} 
  K.~Murase, P.~Meszaros and B.~Zhang,
  Phys.\ Rev.\ D {\bf 79}, 103001 (2009)
  doi:10.1103/PhysRevD.79.103001
  [arXiv:0904.2509 [astro-ph.HE]].

\bibitem{Bartos:2011aa} 
  I.~Bartos, C.~Finley, A.~Corsi and S.~Marka,
  Phys.\ Rev.\ Lett.\  {\bf 107}, 251101 (2011)
  doi:10.1103/PhysRevLett.107.251101
  [arXiv:1108.3001 [astro-ph.HE]].


\bibitem{Abeysekara:2011yu}
  A.~U.~Abeysekara {\it et al.} [HAWC Collaboration],
  Astropart.\ Phys.\  {\bf 35} (2012) 641
  doi:10.1016/j.astropartphys.2012.02.001
  [arXiv:1108.6034 [astro-ph.HE]].


\end{thebibliography}
\end{document}